\documentclass[]{aastex631}

\usepackage{amsmath,amssymb}
\usepackage{hyperref}
\usepackage{multirow}
\usepackage{wrapfig}
\usepackage{mathrsfs}
\usepackage{gensymb}
\usepackage{graphicx}
\usepackage{makecell}
\usepackage{soul} %highlight with \hl{}

\graphicspath{{./}{figures/}}

\begin{document}

\title{Flatfield Calibrations with Astrophysical Sources for the Nancy Grace Roman Space Telescope’s Coronagraph Instrument}

\author[0000-0001-6908-9179]{Erin R. Maier}
\affiliation{Department of Astronomy and Steward Observatory, University of Arizona, 933 N. Cherry Ave., Tucson, AZ 85719, USA}

% please edit as needed
\author[0000-0001-7547-0398]{Robert T. Zellem}
\affiliation{Jet Propulsion Laboratory, California Institute of Technology, 4800 Oak Grove Drive, Pasadena, CA 91109, USA}

\author{M. Mark Colavita}
\affiliation{Jet Propulsion Laboratory, California Institute of Technology, 4800 Oak Grove Drive, Pasadena, CA 91109, USA}

\author[0000-0003-4205-4800]{Bertrand Mennesson}
\affiliation{Jet Propulsion Laboratory, California Institute of Technology, 4800 Oak Grove Drive, Pasadena, CA 91109, USA}

\author[0000-0002-2096-4187]{Bijan Nemati}
\affiliation{Center for Applied Optics, The University of Alabama in Huntsville,
301 Sparkman Drive, Huntsville, AL 35899}
\affiliation{Jet Propulsion Laboratory, California Institute of Technology, 4800 Oak Grove Drive, Pasadena, CA 91109, USA}

\author[0000-0002-5407-2806]{Vanessa P. Bailey}
\affiliation{Jet Propulsion Laboratory, California Institute of Technology, 4800 Oak Grove Drive, Pasadena, CA 91109, USA}

\author{Eric J. Cady}
\affiliation{Jet Propulsion Laboratory, California Institute of Technology, 4800 Oak Grove Drive, Pasadena, CA 91109, USA}

\author{Carey Weisberg}\affiliation{Jet Propulsion Laboratory, California Institute of Technology, 4800 Oak Grove Drive, Pasadena, CA 91109, USA}

\author{Daniel Ryan}
\affiliation{Jet Propulsion Laboratory, California Institute of Technology, 4800 Oak Grove Drive, Pasadena, CA 91109, USA}

\author{Ruslan Belikov}
\affiliation{Ames Research Center, PO Box 1, Moffett Field, CA 94035-1000, USA}

\author[0000-0002-1783-8817]{John Debes}
\affiliation{Space Telescope Science Institute, Steven Muller Building, 3700 San Martin Drive, Baltimore, MD 21218, USA}

\author[0000-0001-8627-0404]{Julien Girard}
\affiliation{Space Telescope Science Institute, Steven Muller Building, 3700 San Martin Drive, Baltimore, MD 21218, USA}

\author[0000-0001-7591-2731]{M. Ygouf}
\affiliation{Jet Propulsion Laboratory, California Institute of Technology, 4800 Oak Grove Drive, Pasadena, CA 91109, USA}

\author[0000-0002-0813-4308]{E. S Douglas}
\affiliation{Department of Astronomy and Steward Observatory, University of Arizona, 933 N. Cherry Ave., Tucson, AZ 85719, USA}

\author[0000-0003-1212-7538]{B. Macintosh}
\affiliation{Stanford University, 382 Via Pueblo Mall, Physics Department, Stanford, California  94305-4060, USA}

\begin{abstract}

The Nancy Grace Roman Space Telescope Coronagraph Instrument is a high-contrast imager, polarimeter, and spectrometer that will enable the study of exoplanets and circumstellar disks at visible wavelengths ($\sim$550--850~nm) at contrasts 2--3 orders of magnitude better than can currently be achieved by ground or space-based direct imaging facilities. To capitalize on this sensitivity, precise flux calibration will be required. The Roman Coronagraph, like other space-based missions, will use on-orbit flatfields to measure and correct for phenomena that impact the measured total effective throughput. However, the Coronagraph does not have internal lamp sources, therefore we have developed a method to perform flatfield calibrations using observations of extended sources, such as Uranus and Neptune, using a combination of rastering the Coronagraph's Fast Steering Mirror, tiling the planet across the field of view, and matched-filter image processing. Here we outline the process and present the results of simulations using images of Uranus and Neptune from the Hubble Space Telescopes Wide Field Camera 3, in filters approximate to the Coronagraph's Band 1 and Band 4. The simulations are performed over the Coronagraph's direct imaging and polarimetric modes. We model throughput effects in 3 different spatial frequency regimes including 1) high spatial frequency detector pixel-to-pixel quantum efficiency variations, 2) medium spatial frequency ``measles'' caused by particle deposition on the detector or other focal-plane optics post-launch, and 3) low spatial frequency detector fringing caused by self-interference due to internal reflections in the detector substrate as well as low spatial frequency vignetting at the edges of the Coronagraph's field of view. We show that Uranus and Neptune can be used as astrophysical flat sources with high precision ($\sim$0.5\% relative error).

\end{abstract}


\section{Introduction}\label{sec:intro}
NASA's next flagship mission, the Nancy Grace Roman Space Telescope (Roman; formerly WFIRST), expected to launch in expected to launch in late 2026 and no later than May 2027, features a 2.4~m primary mirror and two instruments: the Wide Field Imager (WFI) and the Coronagraph Instrument (Coronagraph; formerly abbreviated as CGI). The Roman Coronagraph Instrument is a space-based Technology Demonstration of direct imaging techniques that, by pushing current capabilities to greater, more challenging contrast ratios and smaller inner working angles, will demonstrate key technologies for future missions to directly image temperate, potentially habitable, Earth-like exoplanets. Thus, the Roman Coronagraph will pave the way for next generation direct imaging  \citep{mennesson20}, such as the next large flagship mission recommended for launch in the mid-2040s by the Astro2020 Decadal Survey\footnote{https://www.nationalacademies.org/our-work/decadal-survey-on-astronomy-and-astrophysics-2020-astro2020}.

The Coronagraph includes photometric imaging centered at 575~nm (Band 1; 10\% bandwidth) and 825~nm (Band 4; 12\% bandwidth), polarimetry at these two passbands, and slit spectroscopy with a resolution of R~$\sim$50 centered at 660~nm (Band 2; 17\% bandwidth) and 730~nm (Band 3; 17\% bandwidth). The Coronagraph features the use of an electron multiplying charge-coupled device \citep[EMCCD; e.g.,][]{morrissey18}, a detector that is optimized for low photon count rates (as is the case for very high-contrast observations) by achieving near-zero effective read noise. 

The Roman Coronagraph's technology demonstration Threshold Technical Requirement is: ``Roman shall be able to measure (using the Coronagraph Instrument), with SNR $\geq$ 5, the brightness of an astrophysical point source located between 6 and 9 $\lambda$/D from an adjacent star with a V$_{AB}$ magnitude $\leq$ 5, with a flux ratio $\geq$ 1$\times$10$^{-7}$; the bandpass shall have a central wavelength $\leq$ 600~nm and a bandwidth $\geq$ 10\%.\arcsec\ This requirement is the minimum that the Roman Coronagraph must achieve during its Techology Demonstration; however, as stated above, there are additional ``goal'' observation modes (Band 4 photometry, polarimetry in Bands 1 and 4, and spectroscopy in Bands 2 and 3). In order to achieve and exceed these measurement requirements, Roman Coronagraph Instrument images must be calibrated via various methods such as flatfielding. In the current Roman Coronagraph Instrument calibration budget, flatfields have a notional error allocation of 2\% root-mean-square error (RMSE) on the ratio of the values of any two resolution elements (resels). Assuming that the two resolution elements have the same uncertainty, then each resel has an allocation of $\leq$1.41\% per resel.

Flatfields, in particular, are important as they can correct for spatial sensitivity variations across the field of view at three different scales: high frequencies (e.g., typical detector pixel-to-pixel gain variations), medium frequencies (e.g., ``measles'', which can be caused by particle deposition on the surface of the detector post-launch, as observed with Hubble/WFPC1 \citep[e.g., c.f.,][]{mackenty92}, and low frequencies (e.g., due to fringing in the detector or vignetting by the optics). These three spatial frequencies can vary as a function of the wavelength observed, requiring flatfields to be taken in a similar instrument setup to the observations themselves, as well as over time \citep[as is the case for Hubble/STIS;][]{bohlin99}, requiring flats to be measured as close to the observations as possible. Flats will be taken for the Coronagraph's imaging and polarimetry modes; the spectroscopic modes do not require flatfields as the baseline approach is to use the host star's spectrum to illuminate the same pixels as the spectrum of the planet, thereby dividing out common sources of error, such as those normally addressed with flatfielding.

\begin{figure}[ht]
    \centering
    \includegraphics[width=0.9\textwidth]{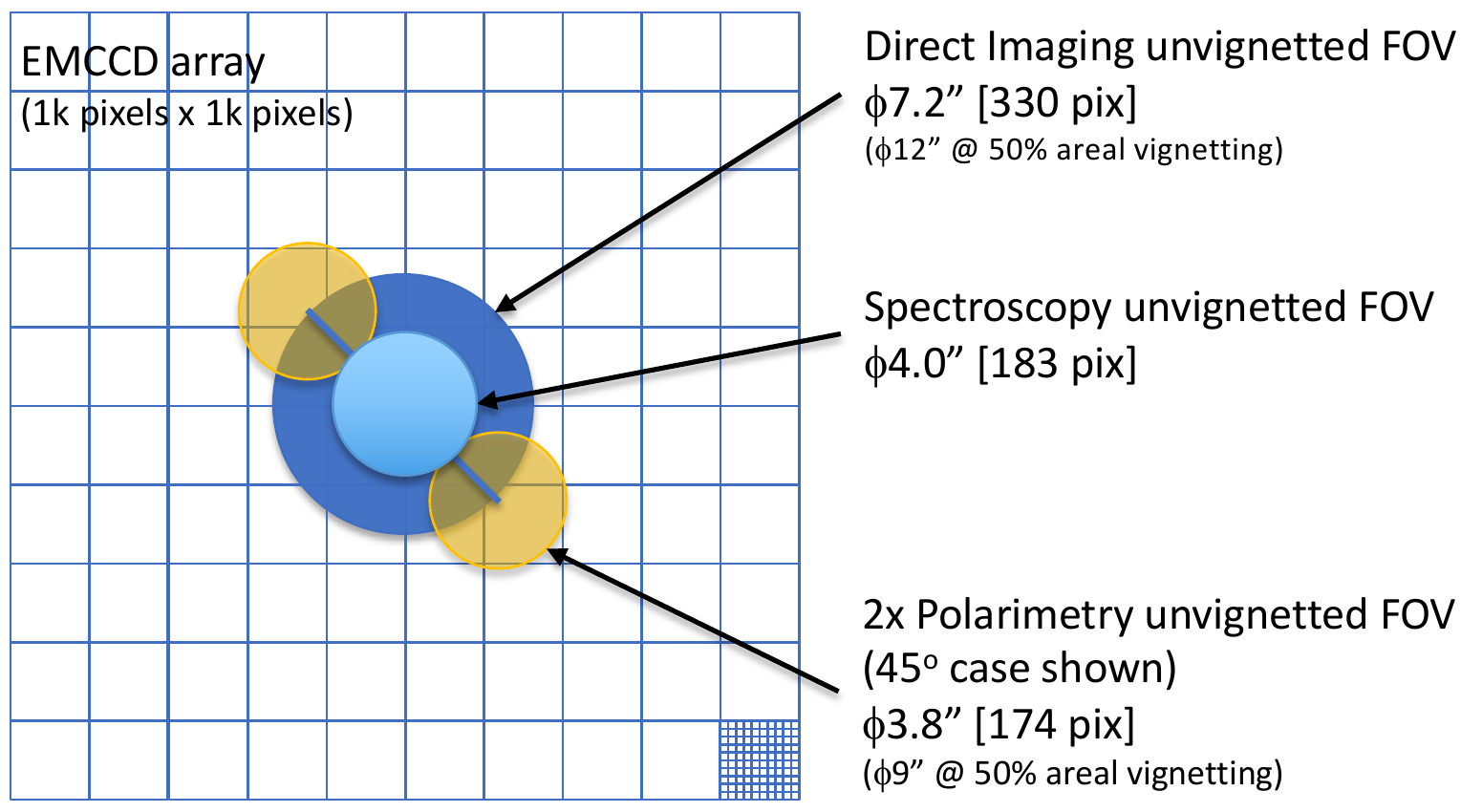}
    \caption{The Roman Coronagraph's various observing modes and associated fields of view: Bands 1 and 4 direct imaging (dark blue), Band 3 spectroscopy (light blue), and Bands 1 and 4 polarimetry (gold). For each Wollaston prism the light is split between the two polarizations (0$^\circ$/90$^\circ$ or 45$^\circ$/135$^\circ$), with the images of the two states separated by 7.5\arcsec\ on the array. Please note that each is the maximum unvignetted FOV for that mode and, in practice, coronagraph mask field stops would decrease the FOV further: 1\arcsec\ diameter for Band 1 with the Hybrid Lyot Coronagraph and 3\arcsec\ diameter for Band 4 Shaped Pupil Coronagraph Wide FOV; see https://roman.ipac.caltech.edu/sims/Param\_db.html\#coronagraph.}
    \label{fig:FOVs}
\end{figure}

Flatfields can be measured via an onboard flat lamp, as is the case with Hubble \citep[e.g.,][]{bohlin99}, or by observing an astrophysical source, such as the zodii with Spitzer \citep{hora08}. Since the Roman Coronagraph Instrument lacks an on-board flat lamp, we present here a detailed study via simulations of using astrophysical sources to sample the unvignetted fields of view (FOV; Bands 1 and 4 direct imaging: 7.2\arcsec\ diameter; Bands 1 and 4 polarimetry: 3.8\arcsec\ diameter; see Figure~\ref{fig:FOVs}) and produce flatfield calibrations.

\section{Astrophysical Objects as Flatfield Sources}\label{Astrophys_objects}

Using astrophysical objects as flatfield sources has heritage from previous space-based observatories. For example, the Cassini Imaging Science Sub-system narrow- and wide-angle cameras used both Venus and Titan as flatfield sources \citep{west10}. Spitzer/IRAC used ``superflats,'' a master flat developed from combining multiple flatfield images of zodiacal light \citep{hora08}. Each flat was produced by dithering over a patch of sky containing zodiacal light, removing all of the stars, and then taking the median value along each pixel. Spitzer/IRAC flatfields were found to be time-invariant over the first two years of operations and therefore all superflats were produced from the first two years of data. Hubble/STIS uses on-board tungsten and deuterium lamps to produce its flatfields \citep{bohlin99}; the flats were found to be time-variant, but independent of wavelength. To be conservative, we adopt the assumption that the Roman Coronagraph's flatfields will also change over time.

In this study, we have primarily explored the use of extended astronomical sources (e.g., a Solar System planet or diffuse extrasolar source such as a galaxy; zodii was not considered in this study due to its relatively dim brightness at the Coronagraph's wavelengths). While we explored using a diffuser illuminated by a star at one of the pupils in the system, given the workable direct imaging procedure described in the following sections, the cost and schedule needed to develop and validate a custom high-performing diffuser could not be justified, and so the diffuser approach was rejected. Using an extended astrophysical source does not require additional hardware, probes the entire optical system, and provides good SNR with modest calibration times.

\subsection{Exploring Extended Astrophysical Sources for Roman}\label{sec:extended_astro_source}

An extended astrophysical source provides the opportunity to illuminate many pixels simultaneously (for reference, the Corongraph's pixel scale is 0.0218~arcsec/pixel), and potentially the entire focal plane, in a single exposure. It has the additional benefit of probing the entire optical train, and not just the Roman Coronagraph back-end. Potential sources considered were Solar System targets, galaxies, and nebulae. Due to Roman's pointing restrictions, the Moon, Earth, Venus, and Mercury cannot be observed as they are too close to the Sun. Due to Roman's lack of non-sidereal tracking and their relatively large proper motions, Jupiter and Saturn ($\lessapprox$0.6\arcsec/minute and $\lessapprox$0.3\arcsec/minute, respectively), as well as their moons, were also eliminated from consideration. These considerations left the relatively-slow-moving Uranus and Neptune ($\lessapprox$0.15\arcsec/minute and $\lessapprox$0.09\arcsec/minute, respectively; see Section~\ref{sec:pointing} for more details) as the only potential Solar System targets for flat sources. Nearby galaxies were also considered as they are large, extended sources that can illuminate the entire focal plane in one exposure. The Andromeda Galaxy (M31) was chosen for further study as it has a mostly spatially-smooth profile, particularly near its center, and is comparatively bright (15.9~V-mag/arcsec$^{2}$). However, we soon found that, despite its brightness, the required exposure times would be prohibitive to the efficient production of flatfields; thus M31 was dismissed. Lastly, nebulae were also briefly considered, but ended up being dismissed for the same reason as galactic targets: they are comparatively dim ($\ge$17~mag/arcsec$^{2}$) over the Roman Coronagraph passbands and would require comparatively long exposure times. Thus, maximizing the Coronagraph's efficiency left us with the aforementioned Solar System targets. In this study we explore in detail Uranus and Neptune, adopting the parameters in Table~\ref{Tab:Planet_Param} and geometric albedos as observed and reported in \citet{karkoschka98}.

\begin{table}[]
\centering
\begin{tabular}{|c|c|c|c|c|}
\hline
\textbf{} & \textbf{V-band} & \textbf{} & \textbf{Surface} & \textbf{Proper} \\
\textbf{Planet} & \textbf{Magnitude} & \textbf{Diameter} & \textbf{Brightness} & \textbf{Motion} \\ \hline

Uranus & 5.68 & 3.4\arcsec & 8.2 V-mag/arcsec$^{2}$ & $\lessapprox$0.15\arcsec/minute \\ \hline
Neptune & 7.7 & 2.3\arcsec & 9.4 V-mag/arcsec$^{2}$ & $\lessapprox$0.09\arcsec/minute \\ \hline
\end{tabular}
\caption{Parameters adopted for Uranus and Neptune for this study. We note that in the case of Uranus, we use the geometric albedos provided in \citet{karkoschka98}, and a 6000~K blackbody. We further adopt Uranus's geometric albedo for Neptune, as its albedo is the same as Uranus's over the Roman Coronagraph bandpasses to first order.}
\label{Tab:Planet_Param}
\end{table}

We explore the availability of Uranus and Neptune for observations with the Roman Coronagraph. We find that although both Uranus and Neptune are visible to Roman for $\sim$6~months of the year, Uranus' observability is severely impacted by Roman/WFI's Galactic Bulge Time Domain Survey and is thus limited to $\sim$2 months of availability. However, Neptune is always accessible when it is visible ($\sim$6 months). Therefore, while Uranus might be an ideal target in terms of its brightness and angular size, Neptune has the benefit of much greater availability.

\subsection{Rastering and Tiling}

Due to their relatively small angular size, neither Uranus (3.4\arcsec\ diameter) nor Neptune (2.3\arcsec\ diameter) completely fill the Coronagraph's full unvignetted field of view. Therefore, we devise a method to dither Uranus and Neptune across the focal plane with multiple telescope pointings (dithers). Such a pattern would illuminate the entire unvignetted focal plane over multiple exposures. While Uranus and Neptune are comparable in size to the baseline Band 1 hybrid Lyot and Band 4 shaped pupil coronagraphic FOVs (0.96\arcsec\ diameter and 2.85\arcsec\ diameter, respectively), requiring fewer dithers, we nonetheless opt to take the conservative approach of requiring flatfielding across the entire unvignetted FOV. The same flatfield approach is applicable to the baseline modes; the number of dithers could simply be reduced when the full unvignetted FOV is not required. Thus, the number of dithers and associated timings presented here are overly-conservative estimates. 

\begin{figure}[ht!]
    \centering
    \includegraphics[width=.9\textwidth]{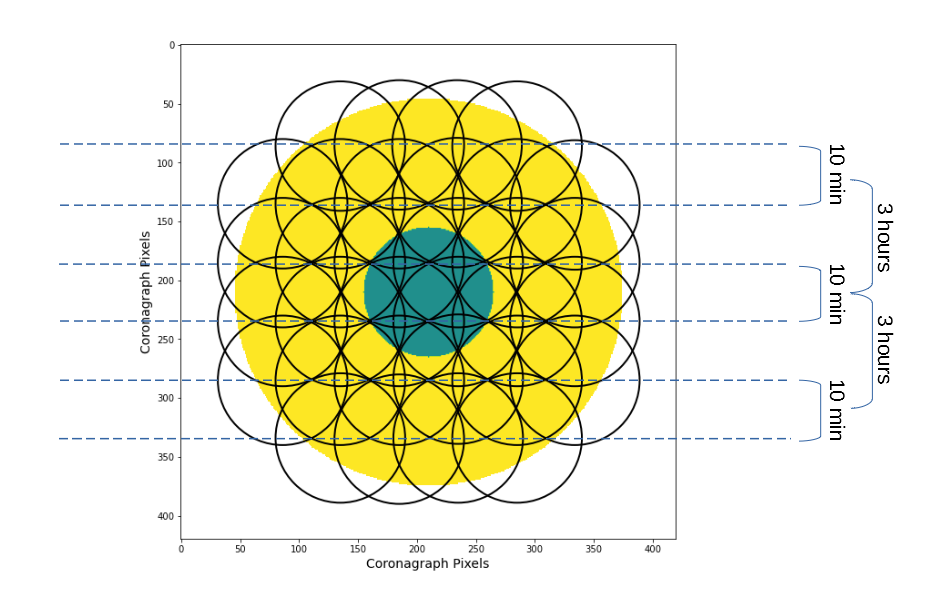}
    \caption{Graphical depiction of the dither tile pattern used to cover the unvignetted focal plane and the time separation of the Hubble source images in our flatfield simulations. The small central solid blue circle is the pre-raster disk of Neptune, the large central solid yellow circle is the Coronagraph's unvignetted FOV, and the black circular outlines show the location of each individual dither tile for Neptune in direct imaging mode. Each row of dithers (horizontal dashed lines) is simulated from the same Hubble/WFC3 UVIS source image. Each pair of rows is simulated from Hubble source images separated by about 10 minutes in time, and each subsequent pair is separated by about 3 hours, for a total timespan of source images of about 6-7 hours over the focal plane. For reference, flats are expected to take $\lessapprox$1~hour for any observing mode, therefore the timespan of the Hubble/WFC3 images is a conservative estimate.}
    \label{fig:tiletime}
\end{figure}

We also adopt a raster by the Coronagraph's fast steering mirror (FSM), a tip/tilt mirror near the entrance to the Roman Coronagraph that can steer the beam, simultaneously during an exposure to further flatten out any spatial variations in the astrophysical source (e.g., due to spots rotating in and out of view on the planet's surface) as well as increase the number of pixels exposed in a single exposure (Figure~\ref{fig:neptconvolve}). We adopt for this initial study a circular raster pattern with a radius of 0.95\arcsec\ conducted over 1~minute; this raster radius is comfortably below the maximum stroke allowed by the FSM, which is is 1.3\arcsec\ in radius. Taking into account this FSM raster and the angular diameters of Uranus and Neptune, we estimate the number of dithers required for the Coronagraph's various fields of view (FOVs) in Table~\ref{tab:inputs}. The overall dither pattern for Neptune in CGI Band 4 can be seen in Figure~\ref{fig:tiletime}. Shown are the size of the disk of Neptune (pre-raster, small solid blue circle); the Coronagraph's unvignetted FOV (large solid yellow circle), and the location of Neptune in each of the individual dither tiles (overlapping black circles). It can be seen that each dither overlaps by about half of the planetary disk. The times listed on the right refer to the Hubble data used in our simulations and are explained further in \S\ref{preparing}.

%\textbf{INSERT PHOTO OF FSM RASTER PATTERN HERE}

\begin{table}[ht!]
\centering

\begin{tabular}{|c|c|c|c|c|c|c|c|}
\hline
\textbf{} & \textbf{} & \textbf{FSM Raster} & \textbf{Dither} & \textbf{Dither Step} & \textbf{Number of} & \textbf{Number of} & \textbf{SNR/resel}\\
\textbf{Planet} & \textbf{Mode} & \textbf{Size (radius)} & \textbf{Step Size} & \textbf{Size} & \textbf{Dithers} & \textbf{Matched Filters} & \textbf{per Dither}\\
\hline
\multirow{2}{*}{Neptune} & Imaging & 0.95\arcsec & 50~pixels & 1.0\arcsec & 32 & 3 & 250 \\ \cline{2-8} 
 & Polarimetry & 0.95\arcsec & 35~pixels & 0.7\arcsec & 12 & 2 & 250 \\ \hline
\multirow{2}{*}{Uranus} & Imaging  & 0.95\arcsec & 55~pixels & 1.1\arcsec & 21 & 2 & 250 \\ \cline{2-8} 
 & Polarimetry & 0.95\arcsec & 40~pixels & 0.8\arcsec & 5 & 1 & 250 \\ \hline
\end{tabular}
\caption{Observing setup for the flatfield process for both planets in both observing modes. Note that the dithers overlap each other by a significant fraction of the planetary radius in both axes, to ensure full coverage and multiple samples of each position; see Figure~\ref{fig:tiletime}.}
\label{tab:inputs}
\end{table}

\subsection{Pointing and Tracking} \label{sec:pointing}

Roman does not support non-sidereal tracking; however, both planets move slowly enough that we can accomplish these observations with a series of sidereal pointings. Uranus has a maximum drift rate during 2026 of $\sim$8.75\arcsec/hour (0.15\arcsec/minute) while Neptune has a maximum drift rate during 2026 of $\sim$5.67\arcsec/hour (0.09\arcsec/minute). Given that each dither will have a total FSM raster time of 1~minute to achieve a total integrated SNR$\geq$250 and the Coronagraph's pixel size of 0.0218\arcsec/pixel, the planets are expected to move 6.69~pixels/minute and 4.33~pixels/minute, respectively; therefore, smearing is a subdominant effect. These shifts are on the order of $\leq$3~resolution elements (``resel''; the half light contour) given the Coronagraph's resel areas of 4.95~pixels$^{2}$ for Band~1 and 9.66~pixels$^{2}$ for Band~4. While the Coronagraph will see some smearing of the targets due to their non-sidereal motion, this motion will further help smooth out any surface features and actually improve the construction of a matched filter. We also added to our total observing time an assumed 30~seconds for the observatory to slew and settle between repointings (dithers). Before beginning a flatfielding sequence, the observatory will be pointed to a nearby star and then point at the planet via absolute-time pointing offsets in right ascension and declination supplied by an ephemeris service such as JPL's Horizons ephemeris service\footnote{https://ssd.jpl.nasa.gov/horizons.cgi} \citep{giorgini96}. The Coronagraph will wait to ``ambush'' the target.

\section{Constructing Roman Coronagraph flatfield Simulations}\label{sec:flatfieldsims}

In this section we describe the data used for the simulations as well as the various effects we model to determine if our on-orbit flatfield procedure can meet our 1.41\%/resel notional allocation. We note that the effects are simulated over a 420x420 Coronagraph pixel region, which is larger than the than the 330 pixel diameter circular unvignetted region, as seen in Figure~\ref{fig:tiletime}. This is to ensure that the dither tiles cover the entire unvignetted region. Data that extends into the vignetted region is not used in the matched filters and statistics are calculated only over the unvignetted region, though the vignetting must still be considered as we will discuss.

\subsection{Preparing the Planet Data}\label{preparing}

\begin{figure}[ht!]
    \centering
    \includegraphics[width=.75\textwidth]{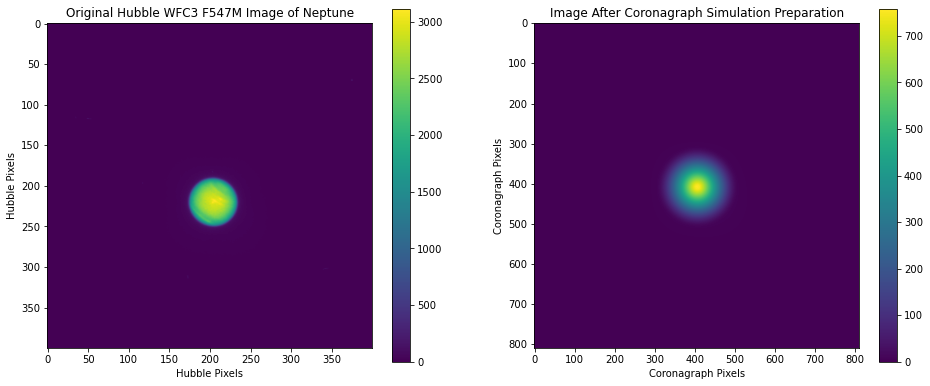}
    \includegraphics[width=.75\textwidth]{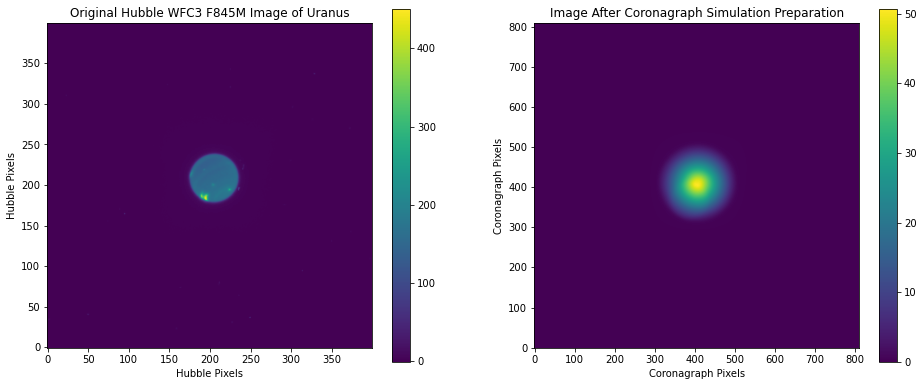}
    \includegraphics[width=.75\textwidth]{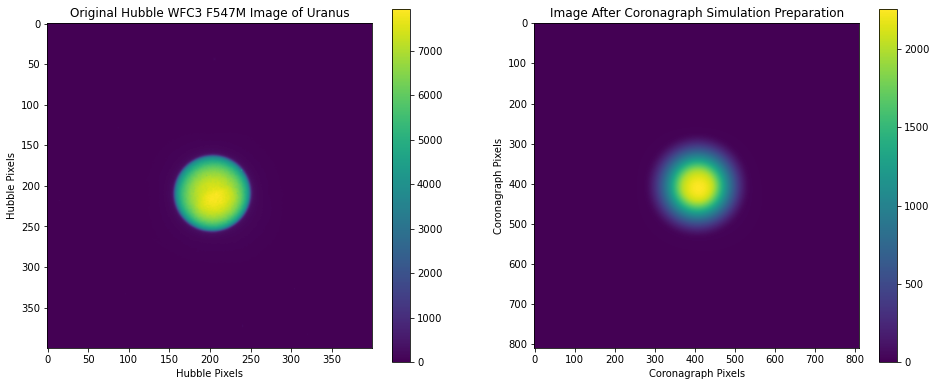}
    \includegraphics[width=.75\textwidth]{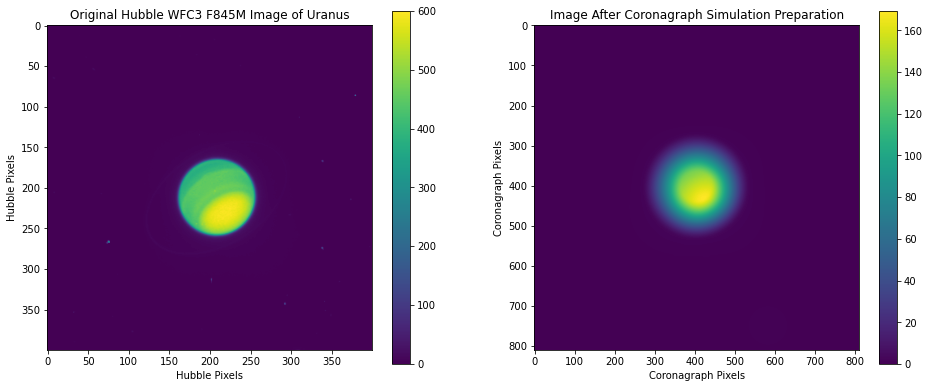}
    
    \caption{$Left$: One each of the Hubble/WFC3 F547M/F845M images of Neptune/Uranus used in the simulations. $Right$: Same images after preparation for injection into the flatfield simulations, including SNR scaling, upsampling to the Coronagraph pixel scale, and convolving with a uniform disk of 0.95\arcsec\ radius to mimic a FSM raster.}
    \label{fig:neptconvolve}
\end{figure}

For these simulations, we utilize Hubble Wide Field Camera 3 (WFC3) images of Uranus and Neptune from the URANUS-MAPS and NEPTUNE-MAPS (Proposal IDs 15262 and 15502, PI: Amy Simon) programs. The images were taken with the WFC3 UVIS imager and F547M (547.5~nm; 13\% bandwidth) and F845M (845.4~nm; 10\% bandwidth) filters, which are similar to the Roman Coronagraph Bands 1 (575~nm; 10\% bandwidth) and 4 (825~nm; 12\% bandwidth), respectively. We use the calibrated data product (\_drz.fits), with bias subtraction, dark subtraction, and flatfielding applied. Figure~\ref{fig:neptconvolve} presents sample Hubble images of Uranus and Neptune in both bands.

We select 6 Hubble images for each planet in each of the two filters for use in our simulations, for a total of 24 different Hubble input images. The separation in time between when the images in each set of 6 were collected by Hubble is shown in Figure~\ref{fig:tiletime}. The first two images were taken 10 minutes apart, then three hours passed, the next two images were taken 10 minutes apart, another 3 hours passed, and then the final two images were taken about 10 minutes apart, for a total time span of 6-7 hours. With our input images selected, we then perform several operations the Hubble data so that it will more accurately represent Coronagraph data. The major factors we consider are plate scale, the Coronagraph FSM raster, and signal-to-noise ratio (SNR).

First we consider the plate scale. We wish to perform our simulations on the Coronagraph pixel scale and then calculate results per resel. As mentioned above, Hubble WFC3 pixels are 40 mas on a side, while Coronagraph pixels are 22 mas on a side, for a 1D scale factor of $\sim1.8$. Thus we remap the data to the Coronagraph pixel scale by upsampling by a factor of 1.8 (without interpolation; we upsample the Hubble of data by a factor of 9 in each axis, then bin by a factor of 5).

 Next, the images are convolved with a 0.95\arcsec\ radius flat disk to simulate rastering during integration by moving the FSM as previously described. We note planet orbital motion (0.15\arcsec/min for Uranus and 0.09\arcsec/min for Neptune) is not included in these simulations, because it is small relative to the diameter of the FSM raster (1.9\arcsec), and exposure times are short ($\sim10$\ s). We also note that we do not simulate individual exposures (i.e. 6 10 second images that are then coadded) but rather simulate one 60 second exposure.

Lastly, we consider SNR. Our simulations assume a final median SNR over the planet disk of 250 per Coronagraph resel. To simulate this, we consider that the Coronagraph has resel areas of $n_{pix}$ = 4.95 pixels$^{2}$\ in Band 1 and $n_{pix}$ = 9.66 pixels$^{2}$\ in Band 4.  We want the median brightness of the planet disk in our upsampled, convolved Hubble images to correspond to an SNR such that $n_{pix}$ pixels of such brightness would have a combined SNR of 250, with the SNR of the rest of the disk scaled relative to this median. To do this, we assign each pixel in the disk in the upsampled, convolved image (Coronagraph pixels) a new value drawn from a random normal Gaussian distribution with a mean of the original pixel value and a standard deviation of

\begin{equation}\label{eq:sigscale}
\sigma = \frac{N_{phot,med}}{250/\sqrt{n_{pix}}} \times \sqrt{\frac{N_{phot,pix}}{N_{phot,med}}}
\end{equation}

where $N_{phot,med}$ is the median number of photons in the planet disk, and $N_{phot,pix}$ is the number of photons in the original individual Coronagraph pixel in the upsampled image. This process is performed on each individual dither tile, so that each image has a unique noise profile, thus allowing us to simulate individual exposures even with the same source image.

These upsampled, convolved, SNR scaled images are the astrophysical input to the simulated flatfield observations. Original Hubble images of Neptune and Uranus in Band 1 and Band 4 and the same images after they have been prepared as described here can be seen in Figure~\ref{fig:neptconvolve}.

\subsection{Detector Vignetting}\label{sec:vignette}

\begin{figure}[ht]
    \centering
    \includegraphics[width=.5\textwidth]{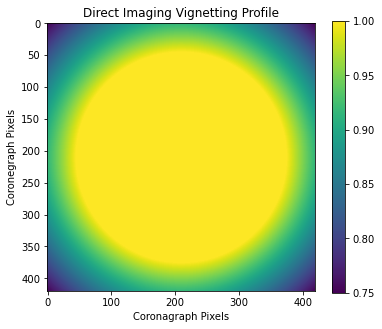}
    \caption{Simulated vignetting profile for direct imaging mode; multiplicative factor is 1 in the 7.2\arcsec\ unvignetted region, linearly decreasing with radial distance to a value of 0.75 at the edge of the simulated DI field. The profile for polarimetric imaging is constructed in the same way but with an minimum value of 0.85.}
    \label{fig:vignette_di}
\end{figure}

While the flatfielding precision notional allocation is specified only for unvignetted portion of the detector, as mentioned above in order to cover this entire region with our flatfielding method we must unavoidably collect some data in the vignetted region. While this data is not used in the matched filters, it must be used as part of the centroiding and cropping steps prior to that (see Sections~\ref{sec:dithrast} and \ref{sec:matchedfilter}), which requires we simulate the vignetting. Based on the areal vignetting estimates in Figure~\ref{fig:FOVs}, we simulate the vignetting profile as a multiplicative factor that is 1 in the unvignetted region, linearly decreasing with radial distance to $\sim$0.75 at the edge of our simulated direct imaging field (Figure~\ref{fig:tiletime}) and $\sim$0.85 for the polarization fields. This simulated profile can be seen for direct imaging in Figure~\ref{fig:vignette_di}.

\subsection{The Roman Coronagraph Spatial and Spectral Fringing Model}

In addition to vignetting, at low spatial frequencies, the detector QE response is affected by detector spatial and spectral fringing.

\begin{figure}[ht!]
    \centering
    \includegraphics[width=.75\textwidth]{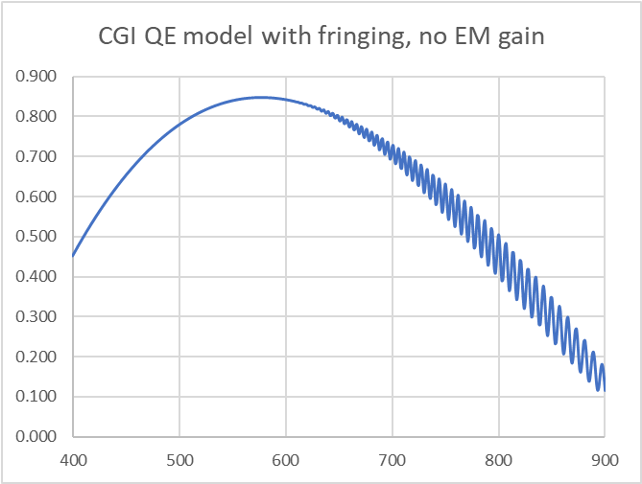}
    \includegraphics[width=.75\textwidth]{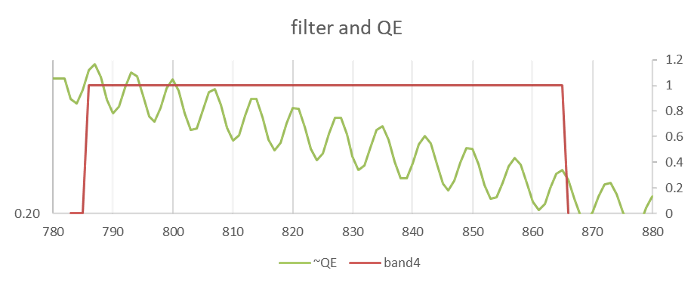}
    \caption{$Top:$ The adopted Roman Coronagraph quantum efficiency model including spectral fringe effects. $Bottom:$ A zoomed-in view of the quantum efficiency and fringing pattern across the Coronagraph's Band 4. Note that spectral fringing amplitude is an order of magnitude larger than the spatial fringing amplitude indicated in Figure~\ref{fig:spatfringing} as the spatial fringing is averaged over the entire the Coronagraph Band~1 passband, which averages down the spectral fringing pattern.}
    \label{fig:spectralfringing}
\end{figure}

To estimate spectral fringing effects in the Coronagraph's EMCCD, we adopt an analytic model for the wavelength dependence of spectral fringing amplitude in the slow f/\# regime; the Coronagraph back end is $\sim$f/50. We started with a nominal Roman Coronagraph EMCCD quantum efficiency (QE) vs. wavelength with no fringing, adopted the fringing amplitude behavior from \citet{groom99}, and then scaled the period for the Coronagraph detector thickness: 12~$\mu$m thick vs. 20~$mu$m for the \citet{groom99} model. We also scaled the starting wavelength of significant fringing of 700~nm seen in \citet{groom99}, blue-shifting it by approximately 80~nm to account for the thinner substrate and the wavelength-dependent Si absorption depth. We show the Coronagraph's estimated spectral fringing model in Figure~\ref{fig:spectralfringing} which indicates that spectral fringe effects are most important at longer wavelengths (e.g., Bands 3 and 4) and are negligible at shorter wavelengths (e.g., Bands 1 and 2). 

The detector fringing can introduce errors if the calibration source spectrum is different from the target spectrum. Assuming that the mean QE spectral slope is known, but not the fringe details, we quantify the errors that arise if one uses Uranus to calibrate, which peaks in the center of Band~4, and then observe a target with a smoother spectrum (modeled here as 4000~K and 10000~K blackbodies). As a function of the fringing phase, the peak relative errors attributable to fringing are 0.4\% with respect to the non-fringing behavior. Our simulations show that for Band 4, using Uranus to calibrate the detector flatfield fringing effects and observing either a 4000 or a 10000~K blackbody target, the estimated maximum flatfielding relative error introduced is 0.24\%. This value is obtained for the worst fringe phase and the worst case number of cycles within Band 4, to allow for uncertainties in exact detector thickness.

\begin{figure}[ht!]
    \centering
    \includegraphics[width=.5\textwidth]{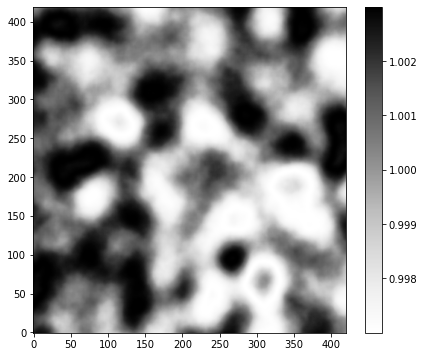}
    \caption{Simulated detector spatial fringing effect on QE in the Coronagraph's Band 4. Note that the fringe effects shown here are smaller in amplitude than the spectral fringing effects illustrated in Figure~\ref{fig:spectralfringing} as the effects demonstrated here are averaged over the Band 4 wavelength range; because the phase of the fringing depends on the wavelength, the pattern is smeared and the amplitude reduced.}
    \label{fig:spatfringing}
\end{figure}

Meanwhile, the map of simulated flatfield contribution due to spatial fringing can be seen in Figure~\ref{fig:spatfringing}. The amplitude of the fringing is matched to the model prediction for Band 4 through the wideband filter. The amplitude through the wideband filter is reduced from the narrowband amplitude shown in Figure~\ref{fig:spectralfringing} due to smearing over the bandpass. The spatial characteristics are notional. In the absence of specific data for the Coronagraph detector array, we use filtered 2D Gaussian noise to roughly match the spatial character of the fringing examples in \citet{groom99}.

\subsection{Measles}

At mid-spatial frequencies, the presence of so-called ``measles," which are artifacts caused by dust, outgassing, and similar effects, manifest as variations in the Coronagraph's observed flatfield. The properties of these measles---size distribution, numbers, distribution of relative opacity variation---on the WFPC1 detector were measured in 1992 are described in detail in the WF/PC Instrument Science Reports (ISRs) 92-8 and 92-11\footnote{https://www.stsci.edu/hst/instrumentation/legacy/wfpc1-instrument-science-reports}. WFPC1 and WFC3 both have pixels sizes of 15 microns, similar to the Coronagraph's pixel size of 13 microns.

The number of measles seen for WFPC1 span about three orders of magnitude, from about $10^{1}$ to $10^{3}$ over a 400x400 WFPC1 pixel region, depending on epoch, morphology, and the detector region considered. For simulation purposes and to remain conservative, we scale the maximum number of measles reported for WFPC1 for any combination of those parameters to the area of the Roman Coronagraph detector being simulated and then double the result, leading to 8000 measles being used in these simulations over a 420x420 Coronagraph pixel area, or about 4000 measles in the unvignetted region.

The measles are simulated as flat disks, with parameters (area, position, and opacity) randomly drawn from distributions seen in the WFPC1 data. Based on the data, the radial size in pixels of each measle is drawn from a normal distribution with a mean of 4 pixels and a standard deviation of 2 pixels. The relative opacity variation is drawn from another random normal distribution with a mean of 0.011 and a standard deviation of 0.013. Finally, the X and Y positions of the center of each measle are drawn from separate uniform distributions with a range of the size of our simulated field, 420 pixels.

\begin{figure}[ht!]
    \centering
    \includegraphics[width=.5\textwidth]{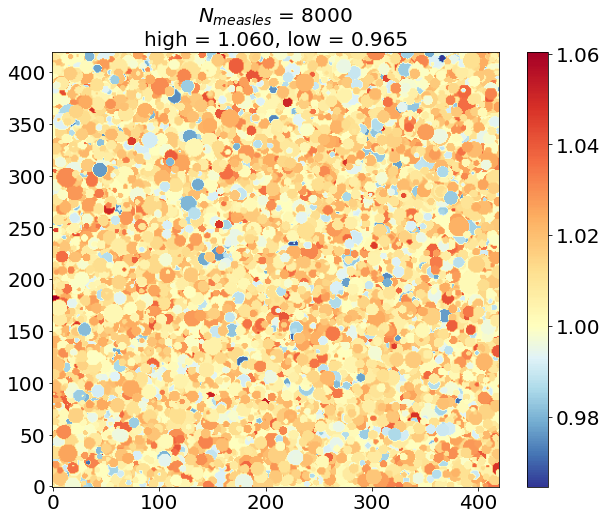}
    \caption{Our simulated measle flatfield mask, containing 8000 measles. This is a little over twice 4000, which is the worst number of measles recorded for 400x400 WFPC1 pixels in 1992, scaled to 420x420 Roman Coronagraph pixels, the size of our simulation region.}
    \label{fig:measles}
\end{figure}

If two measles happen to overlap, the first opacity value is applied to a pixel---so the first measle to fall on that position---takes precedence. The opacity value of the second measle is only applied to pixels that have not previously been affected by a measle. In this way the flatfield mean and standard deviation of the entire focal plane begin to approach the observed distribution parameters in the Hubble WFPC1 data. The final map of simulated flatfield contribution from measles can be seen in Figure~\ref{fig:measles}.

\subsection{Pixel Response Non-Uniformity}

A primary high-spatial frequency effect affecting detector QE response is ``pixel response non-uniformity" (PRNU), an intrinsic variation in the QE response of detector pixels under uniform illumination. The distribution of the PRNU follows an approximately Gaussian distribution with measured response of a 1$\sigma$ value of 3\%, according to EMCCD manufacturing requirements. To simulate this effect, we create a map of relative QE values drawn from a normal distribution with a mean of 1.0 and a standard deviation of 0.03.

\begin{figure}[ht!]
    \centering
    \includegraphics[width=0.5\textwidth]{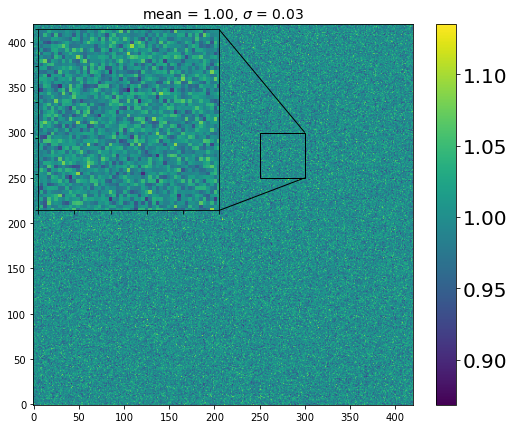}
    \caption{Example Coronagraph pixel-level PRNU mask, with small area blown up to show detail.}
    \label{fig:prnumask}
\end{figure}

\section{Simulating Direct Imaging Flatfield Observations}\label{sec:simdi}

Here we describe and simulate the steps of the flatfield observing sequence. In the following text ``pixel" refers to a Roman Coronagraph pixel. We present figures and results for Neptune in Band 4 in specific, as it is the dimmest out of both planets and bands and thus expected to produce the least accurate (relatively) flatfield.

\subsection{Dithering and Rastering}\label{sec:dithrast}

As previously mentioned we must dither the planet over the unvignetted focal plane and stitch these dithers together to create our flatfield images. 

We start by first creating the total simulated flatfield mask by multiplying the measles, PRNU, spatial fringing, and vignetting masks (created as described in \S\ref{sec:flatfieldsims}) together. We then simulate our individual dither images by multiplying the flatfield mask by the input planet image, moving the mask in X and Y in increments of the chosen dither step size to ensure that every pixel of the flatfield mask is sampled by the planet disk. The dither step size is chosen based on the size of the Hubble planet data arrays and the number of dithers needed to cover the full unvignetted focal plane with minimal extension into the vignetted region, while staying within the notional error allocation. Dither and raster parameters used in these simulations are presented in Table~\ref{tab:inputs}.

Each row of dithers, as seen in Figure~\ref{fig:tiletime} is made from a different one of the 6 input images for the given planet/band, repeated multiple times (6 for Neptune for direct imaging; 5 for Uranus) to cover the width of the field. By using different images at different locations in the focal plane, we are able to simulate having taken multiple exposures over a period of time, including the effects of planet rotation. However, while we simulate a square number of images for ease of programming, we then remove the 4 corner tiles, as they extend the furthest into the vignetted region. Operationally, these 4 images would just not be collected. 

An additional point of concern here is Roman's pointing stability, which impacts our ability to place the planet at the desired position in the focal plane, and thus may impact our results. Roman's pointing stability is assumed at 0.3 arcsec / 3s 1-$\sigma$ for initial pointing and 8$\sim$mas in jitter. While a robust simulation of sensitivity to pointing stability is left to future work, Uranus and Neptune are only required to expose the entire unvignetted portion of the focal plane over multiple dithers, so some slack due to pointing uncertainties can be tolerated as long as we can accurately centroid on the planet to construct a matched filter.

We perform an analysis in which we compared the known centroid of the input Hubble/WFC3 image to the measured centroid in the simulated CGI image. We compare the actual values to the predicted values by taking the standard deviation of the predicted value minus the calculated value in each axis. For Neptune in Band 4 we find that the standard deviation is less than 1 pixel in both axes. These results indicate that we can centroid very well on the planet when creating flatfield images. Thus even if the planet is off the desired position for a given dither due to pointing instability, we will be able to accurately calculate the position and thus accurately stack the dithers to create our matched filters as described in \S\ref{sec:matchedfilter}.

\subsection{Creating the Matched Filters}\label{sec:matchedfilter}

\begin{figure}[h!]
    \centering
    \includegraphics[width=\textwidth]{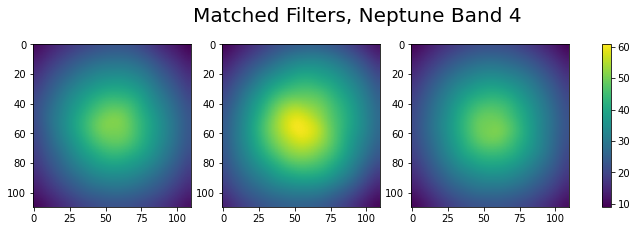}
    \caption{The 3 matched filters created for Neptune in Band 4 made from 10, 12, and 10 of our dither images.}
    \label{fig:matchedfiltersnept}
\end{figure}

Next we create matched filters of the planet to solve for the common-mode astrophysical signal. First, we crop each individual dither image, centered on the planet, to a radius slightly less than the planet radius to minimize the effects of limb darkening, which comparatively degrades the SNR for those pixels. Next, we assign all pixels in each dither image outside of the unvignetted region to the value of NaN, as we do not want to include pixels that are vignetted in our matched filters. Each cropped image is then smoothed again (to be clear, a separate step from the raster) with a 3$\sigma$ (0.06\arcsec) Gaussian kernel to minimize photon-noise errors. This kernel size struck the balance between smoothing out the images too much vs. too little, such that flatfield retrieval could be achieved with the required precision. We then median combine subsets of the dither images to produce the desired number of matched filters. For the shown simulations of Neptune in Band 4 we create 3; one for the first 10, middle 12, and last 10 images, seen in Figure~\ref{fig:matchedfiltersnept}. The number of matched filters for the other modes are presented in Table~\ref{tab:inputs}.

\subsection{Calculate Simulated Flatfield Mask}

Now we divide the dither images by their corresponding matched filters. The matched filters effectively remove the common-mode astrophysical signal, leaving us a residual that is the section of flatfield mask originally applied to each region of the detector.

Each of these residual fields is then cropped to a disk shape with a radius 5 pixels less than the original crop radius, as using the full radius leads to edge effects when stitching the fields together. All of the residual fields are then stitched together into our final retrieved flatfield mask, with the average value taken in the places where the fields overlap.

We now want to calculate the precision of our flatfield retrieval at the resel level. To do this, as the Coronagraph resels are defined as the aperture that would be used in that band for aperture photometry, we convolve both of the original and retrieved flatfield masks with a circular aperture with the pixel radius of the resel in the given band (1.26 pixels Band 1; 1.75 pixels Band 2). We then calculate the percent difference between original and retrieve and then calculate the standard deviation of the percent difference over the unvignetted focal plane. We then add potential error from spectral fringing (Band 1: $\sim$0\%, Band 4: 0.24\%), via a root-sum-square with the standard deviation of the percent difference. This value is the final accuracy of the retrieval of the original flatfield mask, to be compared to the notional allocation of 1.41\%/resel.

\begin{figure}[h!]
    \centering
    \includegraphics[width=\textwidth]{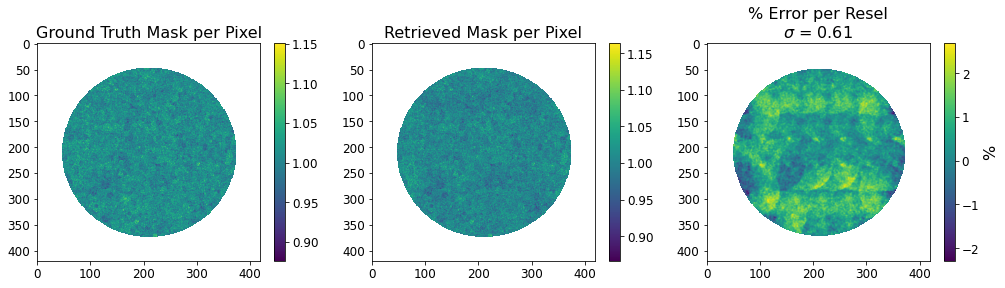}
    \caption{The ground truth flatfield mask, recovered flatfield mask, and \% difference with standard deviation per resel for the entire unvignetted focal plane for Neptune in Band 4. This does not include the spectral fringing error component. Axis units are in Coronagraph pixels.}
    \label{fig:recoverytriplet_full}
\end{figure}

Figure~\ref{fig:recoverytriplet_full} presents the original mask, retrieved mask, and retrieval accuracy for the entire unvignetted focal plane for Neptune in Band 4 using 3 different matched filters, with an overall accuracy of 0.61\%, which increases to 0.65\% when the spectral fringing term of 0.24\% is added. We find that with more matched filters, accuracy is dominated by the SNR of each planet image. With fewer matched filters, it is dominated by the variability of the planet. The repeated structures and gradients visible in the percent different plot are due to matched filter mismatch due to centroiding errors and seem to constitute the majority of the retrieval error. That some data originally extended into the vignetted region, affecting the centroiding, also appears to contribute about 0.1-0.2\% to the error. Nevertheless, we have shown that for Neptune in Band 4, expected to have the least accurate flatfield retrieval, our accuracy of 0.65\% is well with in the direct imaging notional allocation of 1.41\%. The results for both planets in both bands can be seen in Table~\ref{tab:di_results}.

\section{Simulating Polarimetry Observations}

We now proceed to determine if the dither and raster flatfielding procedure can be done to the notional accuracy of 2.5\% per resel for the Coronagraph's polarization mode. This comparatively relaxed allocation is due to the polarimetric observing strategy, which will observe multiple polarization standards for calibration at the same location on the detector. The polarized images for the Coronagraph are generated with a Wollaston prism. Each Wollaston prism creates two 3.8\arcsec\ diameter images simultaneously, separated by 7.5\arcsec: one for the 0$\degree$ polarization and one for the 90$\degree$ polarization (or $45\degree$ and $135\degree$). The unvignetted size of each of these 2 polarization focal planes is 174 Coronagraph pixels in diameter. We only simulate here the case of the 2 images obtained at linear polarizations of 0$\degree$ and 90$\degree$, as the results obtained for 45$\degree$ and 135$\degree$ images should be similar. 

Uranus and Neptune are both slightly linearly polarized, as shown by previous observations obtained with the ESO 3.7m La Silla telescope, and their degree of polarization varies both spatially - from the center of the planet to the limb - and as a function of wavelength \cite{joos07}. \citet{Schmid2006} observed Neptune in R-band (Bessell; $\lambda_{c,eff}=786$~nm, $\Delta\lambda_{eff}=140$~nm) and z-band (Gunn; $\lambda_{c,eff}=890$~nm, $\Delta\lambda_{eff}=100$~nm).

In order to simulate the spatial variations of polarization across the planet disk, we use the same r(x,y) quantity measured by \citet{Schmid2006} (their Figures 2 and 3), which is defined as 
\begin{equation}\label{eq:rxy1}
    r(x,y) = \frac{I0(x,y) - I90(x,y)}{I_{peak}}
\end{equation}
where $I_{peak}$ is the maximum unpolarized intensity measured on the planetary disk (Hubble/WFPC3 data).

\begin{figure}[ht!]
    \centering
    \includegraphics[width=\textwidth]{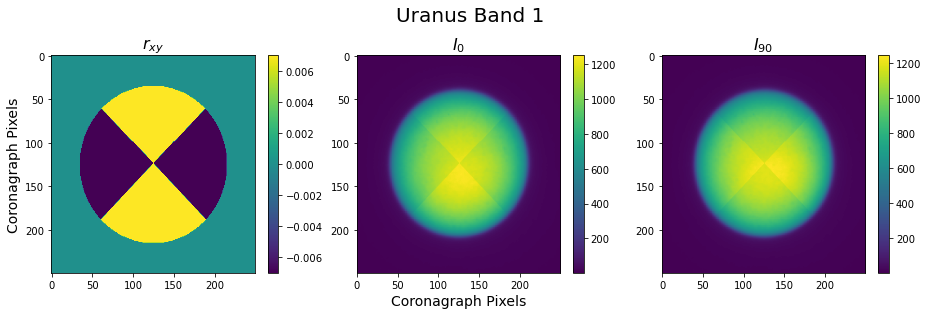}
    \caption{The injected r(x,y) intensity pattern, approximated from Figure 3 in \citet{Schmid2006}, with a magnitude of $\pm$1.15\%. for Uranus in Band 1. The second and third panels show the simulated I0 and I90 images on the Coronagraph pixel level, with the injected pattern visible.}
    \label{fig:rxypaper}
\end{figure}

In order to simply represent the spatial variations measured - and maximize their impact on polarimetric flat fielding errors - we simulated a simple binary r(x,y) map (Figure~\ref{fig:rxypaper}, left panel). It contains four quadrants with regions alternating between r(x,y) = -0.005 and +0.005 for Uranus and Neptune in Band 4 (0.005 corresponds to the maximum spatial variation of r observed over the disk), between -0.0115 and +0.0115 for Uranus in Band 1, and between -0.006 and +0.006 for Neptune in Band 1.

A comparison of the r(x,y) values reported in \citet{Schmid2006}, and the standard Stokes (Q/I)(x,y)
\begin{equation}\label{eq:stokes}
    (Q/I)(x,y) = \frac{I0(x,y) - I90(x,y)}{I0(x,y) + I90(x,y)}
\end{equation}
value reported by \citet{joos07} indicate that r(x,y) \~ 0.5 (Q/I) (x,y).

Based on this, given our unpolarized Hubble/WFC3 images $I_{unpolarized}$(x,y), after resampling to the Coronagraph pixel level and SNR scaling, we then use our simulated r(x,y) maps to generate the ground truth simulated images for 0 and 90 degree polarization for each planet (Figure~\ref{fig:rxypaper}, middle and right panels), henceforth I0 and I90, as

\begin{equation}\label{eq:i0fin}
    I0 = 0.5 I_{unpolarized}  (1 + (2*r(x,y)))
\end{equation}

\begin{equation}\label{eq:i90fin}
    I90 = 0.5 I_{unpolarized}  (1 - (2*r(x,y)))
\end{equation}

We show results of our simulations for Uranus in Band 1 here as it is expected to have the most significant polarization fraction. As in direct imaging, we convolve our initial I0 and I90 images with a flat disk of radius 0.95\arcsec\ to simulate the FSM raster. We then create two 250x250 Coronagraph pixel diameter flatfield masks, one for I0 and one for I90, with independent PRNU, measle (3000 measles for this area), and fringing realizations (given that the I0 and I90 images cover distinct regions of the detector), and the vignetting profile described in Section \ref{sec:vignette} is applied to both.  We assume that the statistical properties of all flatfield effects are polarization-independent. We then follow the same process as in direct imaging to retrieve the estimated flatfield mask and compare it to the ground truth, using the input parameters in Table \ref{tab:inputs}.

\begin{figure}[ht!]
    \centering
    \includegraphics[width=\textwidth]{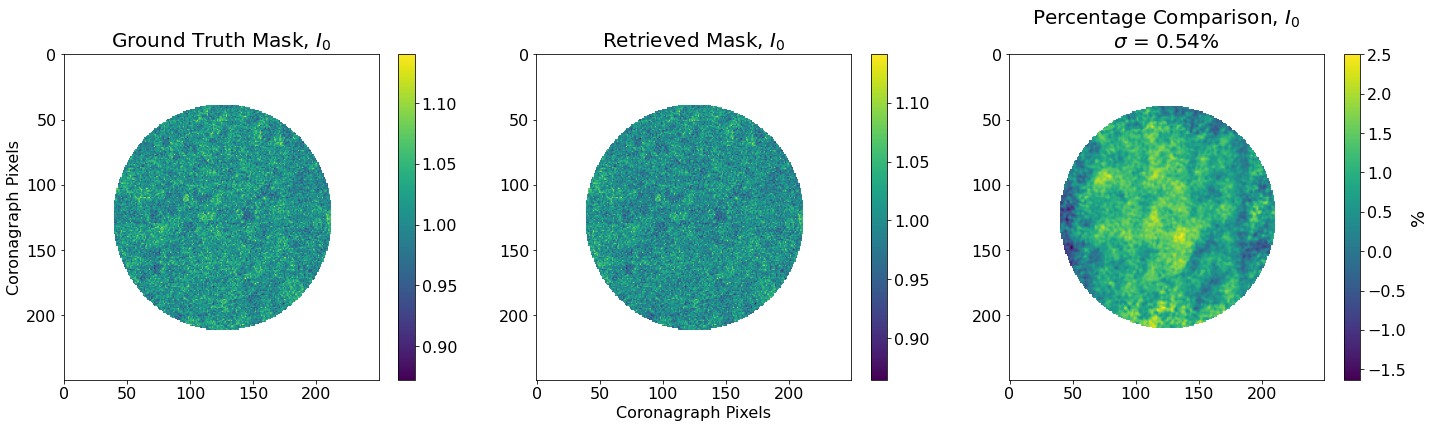}
    \caption{The ground truth flatfield (1st panel) and retrieved flatfield mask (second panel) for the I0 field at the pixel level, and the percent difference also at the pixel level with $\sigma$ at the resel level (third panel).}
    \label{fig:fullmask_0}
\end{figure}

\begin{figure}[ht!]
    \centering
    \includegraphics[width=\textwidth]{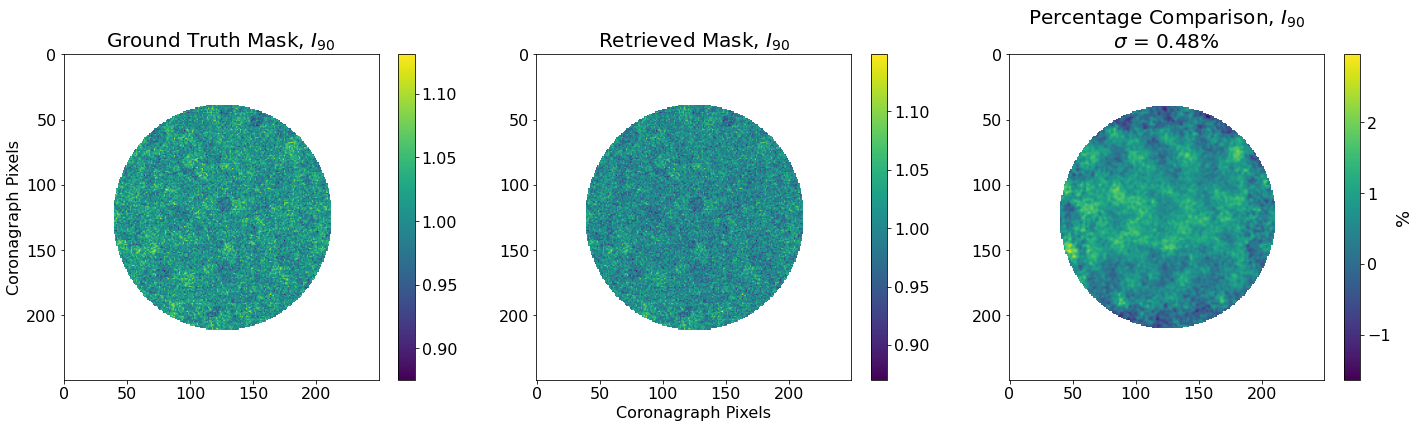}
    \caption{The ground truth flatfield (1st panel) and retrieved flatfield mask (second panel) for the I90 field at the pixel level, and the percent difference also at the pixel level with $\sigma$ at the resel level (third panel).}
    \label{fig:fullmask_90}
\end{figure}

The full mask retrievals for both polarization fields and their percent error per resel for Uranus in Band 1 can be seen in Figures \ref{fig:fullmask_0} and \ref{fig:fullmask_90}. We see that again the errors are well within the notional allocation of 2.5\% for the polarimetric mode. The results per planet, band, and polarization can be seen in Table~\ref{tab:pol-results}. 

\begin{figure}[ht!]
    \centering
    \includegraphics[width=.6\linewidth]{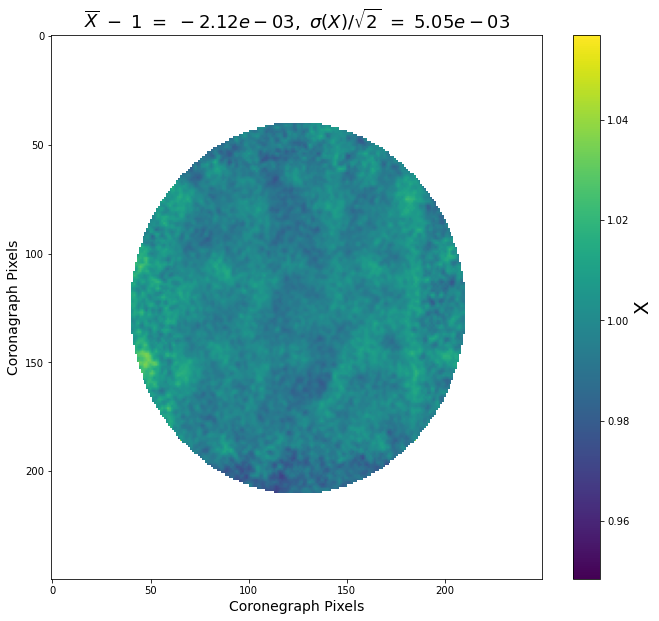}
    \caption{X (Equation \ref{eq:xratio}) for Uranus in Band 1. The reported statistics show that there is not a systemic bias between the two different detector regions for polarimetic imaging ($\overline{X} - 1 \sim 0$), and that there is not a spatial correlation of errors between the two region ($\sigma(X)/\sqrt{2} \sim \sigma(0) \sim \sigma(90)$}
    \label{fig:xratio}
\end{figure}

 In addition, we then calculate the quantities $X_{0}(x,y)$ and $X_{90}(x,y)$, which are the ratio of the retrieved flatfield mask to the ground truth flatfield mask for each polarization field. We then calculate 
 
 \begin{equation}\label{eq:xratio}
     X = \frac{X_{0}(x,y)}{X_{90}(x,y)} = \frac{Mask_{0,est}\ /\ Mask_{0,ground}}{Mask_{90,est}\ /\ Mask_{90,ground}}
 \end{equation}\newline
 
 which can be seen in (Figure~\ref{fig:xratio}). We look at this ratio because the 0 and 90 polarimetric images are formed on different parts of the detector. X is sensitive to any average flatfielding error, i.e., systemic bias, as well as any spatially correlated error between the two regions. To test for systemic bias, we calculate $\overline{X} - 1$, which should be around 0 as X should be approximately flat around a mean of 1. To test for spatial correlation of errors, we calculate $\sigma(X)/\sqrt{2}$, which in the case of no spatial correlation and approximately equal flatfielding error of value $\sigma$ in the 2 region should be equal to $\sigma$. In the case of Uranus in Band 1 in Figure~\ref{fig:xratio}, we see that $\overline{X} - 1$ is indeed close to 0, and that $\sigma(X)/\sqrt{2}$ is approximately equal to the average of the 0 and 90 flatfielding errors, both as expected. Similar results were obtained for both planets in both bands.

\begin{figure}[ht!]
    \centering
    \begin{minipage}{.45\textwidth}
    \includegraphics[,width=\textwidth]{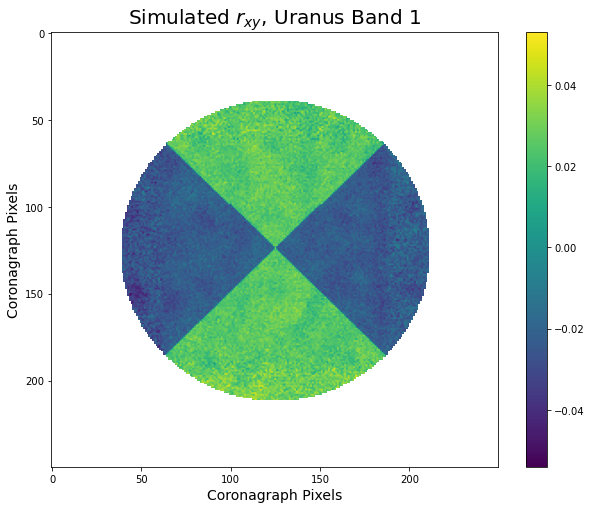}
    \end{minipage}
    \begin{minipage}{.45\textwidth}
    \includegraphics[width=\linewidth]{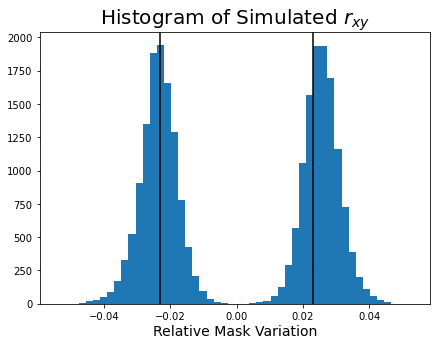}
    \end{minipage}
    \caption{Left: Simulated r(x,y) per Coronagraph pixel made from the simulated I0 and I90 images which have been flattened using the retrieved flatfield masks. Right: Histogram of the simulated r(x,y). Magnitude should be approximately twice the injected r(x,y), or about 2.3\% for Uranus in Band 1, which it is (vertical lines mark $\pm$2.3\%).}
    \label{fig:rxycompare}
\end{figure}

We also test our ability to retrieve the original injected r(x,y). To do this we calculate simulated flatfielded I0 and I90 images by multiplying the upsampled/SNR scaled but unconvolved Hubble images by their respective ground truth flatfield masks, then divide them by their retrieved flatfield masks. Then we calculate r(x,y) per Coronagraph pixel as in Equation \ref{eq:rxy1}. The simulated r(x,y) can be seen in Figure~\ref{fig:rxycompare}. The original binary quadrant pattern is present. Based on Equations \ref{eq:rxy1}-\ref{eq:i90fin} we should expect our retrieved r(x,y) to have twice the magnitude of the injected r(x,y), or about 2.3\% for Uranus in Band 1, which the histogram shows to be correct (vertical lines mark $\pm$2.3\%).

\section{Summary of Results}

In Tables~\ref{tab:di_results}--\ref{tab:flattimes} we present a summary of our analysis of the dither and raster method for producing accurate flatfields for the Coronagraph for both planets (Uranus and Neptune), bands (Band 1 and 4), and modes (direct imaging photometry and polarization). Tables~\ref{tab:di_results} and \ref{tab:pol-results} show the precision of the flatfield mask retrieval per resel for direct imaging photometry mode and polarimetry mode, respectively, including the spectral fringing error via RMS for the Band 4 results.

There is generally some variation in the precision results with different realizations of the simulated flatfield effects, which appears to be dominated by centroiding errors and error due to mismatch of the matched filters. However, this variation is typically less than $\sim$0.1\%; still well within the error allocations. As previously noted, the notional allocation per resel for direct imaging photometry is 1.41\%, and for polarimetry is 2.5\%, and the raster and dither method clearly meets these allocations with ample margin for both modes, for both planets, bands, and polarizations.

The time required to produce the Coronagraph flats for all observing modes using Neptune and Uranus as flatfield sources is shown in Table~\ref{tab:flattimes}. While Uranus would be the preferred source as it only takes 2.4~hours to construct a flatfield (due to its comparatively larger angular size and surface brightness), it is less visible over the course of a year compared to Neptune due to the timing of the Roman/WFI's Galactic Bulge Time Domain Survey.

\begin{table}[ht!]
\centering
\begin{tabular}{|c|c|c|}
\hline
\textbf{Planet} & \textbf{Band 1} & \textbf{Band 4} \\ \hline
Neptune & 0.53\% & 0.65\% \\ \hline
Uranus & 0.45\% & 0.58\% \\ \hline
\end{tabular}
\caption{Standard deviation of relative flatfield estimation error per resel across entire unvignetted focal plane (7.2\arcsec\ diameter), per planet and band, in direct imaging photometric mode. For Band 4, the 0.24\% error from spectral fringing is included through a root-sum-square.}
\label{tab:di_results}
\end{table}

\begin{table}[ht!]
\centering
\begin{tabular}{|c|c|c|c|c|c|c|c|c|}
\hline
\multirow{2}{*}{\textbf{\begin{tabular}[c]{@{}c@{}}Polarization Angle\end{tabular}}} & \multicolumn{4}{c|}{\textbf{Neptune}} & \multicolumn{4}{c|}{\textbf{Uranus}} \\ \cline{2-9} 
 & \multicolumn{2}{c|}{\textbf{Band 1}} & \multicolumn{2}{c|}{\textbf{Band 4}} & \multicolumn{2}{c|}{\textbf{Band 1}} & \multicolumn{2}{c|}{\textbf{Band 4}} \\ \hline
0$\degree$ & \multicolumn{2}{c|}{0.57\%} & \multicolumn{2}{c|}{0.55\%} & \multicolumn{2}{c|}{0.54\%} & \multicolumn{2}{c|}{0.55\%} \\ \hline
90$\degree$ & \multicolumn{2}{c|}{0.47\%} & \multicolumn{2}{c|}{0.58\%} & \multicolumn{2}{c|}{0.48\%} & \multicolumn{2}{c|}{0.55\%} \\ \hline
\end{tabular}
\caption{Standard deviation of relative flatfield estimation error per resel (in percent) across the entire unvignetted focal plane (3.8\arcsec), per planet, polarization state, and band, in polarimetry mode. Spectral fringing error 0.24\% is included for Band 4 numbers in root-mean-square. Note that in polarimetry mode, images from orthogonal polarization states are recorded on different detector regions, and both regions are independently flatfielded within the required 2.5\% rms per resel precision. }
\label{tab:pol-results}
\end{table}

\begin{table}[ht!]
\centering
\begin{tabular}{|l|l|l|}
\hline
& \textbf{Neptune} & \textbf{Uranus}\\ \hline
Total Integration Time (Bands 1 and 4)  & {5.1 hours}            & {1.9 hours} \\
\hline
Total Overhead Time (@30~sec/tile) & {0.9 hours}            & {0.6 hours} \\
\hline
Total Calibration Time & {6.0 hours}            & {2.5 hours} \\
\hline
\end{tabular}
\caption{Time to Produce Flats (Bands 1 and 4 direct imaging and polarimetry) - Current Best Estimate; Roman Coronagraph End of Life. Note that all individual integration times in the total were rounded up the next higher minute to match the 1 minute FSM raster period.}
\label{tab:flattimes}
\end{table}

\section{Conclusions and Future work}
The Nancy Grace Roman Space Telescope's Coronagraph Instrument will provide the necessary demonstration of direct imaging technology for the next generation space telescopes, such as a mission that would launch in the mid-2040s to search for biosignatures, as recommended by the Astro2020 Decadal Survey. To help achieve optical operations, calibration datasets, such as flatfields, will be applied to the Coronagraph's images. However, the Roman Coronagraph lacks an on-board flatfield lamp. We have explored here the ability to use astronomical objects as flat sources by tiling images of the object across the unvignetted Coronagraph detector area while simultaneously performing a raster with the FSM to smooth out variations, and then combining these images into one or more matched filters that are used to divide out the common astrophysical signal in the images, leaving us with the flatfield calibration image for that area of the detector. 

We find that Neptune and Uranus can be used with this technique to produce high-precision flatfields ($\sim$0.5\%/resel relative error) for the Corongraph's Bands 1 and Bands 4, in both polarimetric and direct imaging photometry modes, with ample margin on the notional error allocations (1.41\% direct imaging, 2.5\% polarimetry), in a modest amount of time ($\leq$6.0 hours). 

Future work will consider several points which we have highlighted here, including an explicit simulation of the raster pattern as opposed to approximating it with a uniform disk. We also will include Roman pointing instability and planet orbital motion, and perform more robust modeling of the required number/timespan of images included in each matched filter.

\section*{Acknowledgements}
Part of the research was carried out at the Jet Propulsion Laboratory, California Institute of Technology, under contract with the National Aeronautics and Space Administration. Copyright 2022. All rights reserved.

The authors would like to thank Bala Balasubramanian for their help with in-lab testing and comments on this paper and Natasha Batalha for supplying albedo models. The authors would also like to thank Nikole Lewis, Tyler Groff, Patrick Lowrance, Tiffany Meshkat, and Mark Swain for their helpful discussions.

This work made use of various Python packages for computing and plotting, including Matplotlib \cite{Hunter07}, NumPy \cite{Oliphant06}, IPython \cite{Perez07}, Jupyter notebooks \cite{Kluyver16a}, SciPy \cite{2020SciPy-NMeth}, AstroPy\cite{astropy:2013,astropy:2018}, scikit-image\cite{scikit-image}, and photutils\cite{larry_bradley_2020_4044744}.

\bibliography{flatfield}{}
\bibliographystyle{aasjournal}

\end{document}